# LOW-MASS ( M<1.2 GeV/c² ) $\sigma_0$ - MESON PRODUCED IN THE SYSTEM $\pi^+\pi^-$ FROM THE REACTION np → np$\pi^+\pi^-$ AT P$_n$=5.20 GeV/c

Yu.A.Troyan[1*], A.V.Beljaev[1], A.Yu.Troyan[1], E.B.Plekhanov[1], A.P.Jerusalimov[1], S.G.Arakelian[2]

[1]- Veksler and Baldin Laboratory of High Energies, JINR, Dubna.
[2]- Lebedev Institute of Physics, Russian Academy of Sciences, Moscow
*-E-mail: atroyan@jinr.ru

## Abstract

This work continues a series of our publications dedicated to a search and study of low-mass ($M < 1.2$ GeV/c²) $\sigma_0$ - mesons produced in reaction $np \to np\pi^+\pi^-$ at P$_n$ = (5.20 ± 0.12) GeV/c.

In comparison with the above mentioned publications, more strict criteria have been used by us in this work to select events on track measurements accuracy as well as on the event selection of the reaction under consideration.

The spectrum of effective masses of $\pi^+\pi^-$-combinations was studied for the events in which a secondary proton flies forward to the centre of mass system of the reaction ($cos\Theta_p^* > 0$). This kind of selection enabled us to reduce the background from $\pi$-mesons belonging to the nucleon isobars.

In the distribution of the $\pi^+\pi^-$ combinations effective masses we have found 8 peculiarities.

All the observable peculiarities have spins equal to zero.

We have also investigated studied systems $\pi^-\pi^-$ and $\pi^-\pi^0$ in the reactions $np \to pp\pi^+\pi^-\pi^-$ and $np \to pp\pi^-\pi^0$. There were no picks corresponding to those found in the reactions $np \to np\pi^+\pi^-$ in the distributions of the effective masses of the combinations mentioned above. Thus, it follows that isotropic spins of all the peculiarities observed by us in the system $\pi^+\pi^-$, are equal to zero.

So, we have observed a row of resonances with a set of quantum numbers $0^+(0^{++})$ - $\sigma_0$ -mesons decaying on the channel $\sigma_0 \to \pi^+\pi^-$.

## 1. Introduction

This work is devoted to a search for and study of low-mass (M < 1.2 GeV/c²) resonances in the $\pi^+\pi^-$-system. Their existence can clarify the properties of low-lying scalar mesons (the so-called $\sigma_0$-mesons), whose investigation is important both for the mechanism of realization of chiral symmetry for corresponding Lagrangians and for an adequate description of an attractive part of the nucleon-nucleon interaction potential [1].

Different theoretical models give various predictions for masses and widths of $\sigma_0$-mesons. Early quark bag models gave $M_{\sigma_0} > 1.5$ GeV/c² and $\Gamma_{\sigma_0} \geq 0.5$ GeV/c² [2]. Later works predicted $M_{\sigma_0} = 500 \div 1000$ MeV/c² and $\Gamma_{\sigma_0} = 200 \div 500$ MeV/c² for low-lying $(q\bar{q})$-states [3]. Some models of a spontaneous break of chiral symmetry predict $M_{\sigma_0} \approx 700$ MeV/c² and $\Gamma_{\sigma_0} \geq 500$ MeV/c² [4]. Using QCD sum rules and assuming that the $\sigma_0$-meson is a low-lying glueball, the calculations give the following predictions: $M_{\sigma_0} = 280 \div 700$ MeV/c² and $\Gamma_{\sigma_0} = 2 \div 60$ MeV/c² [5] (see also [6]).

As one can see from the works mentioned above there are no common understanding of nature of $\sigma_0$- mesons and therefore its properties. This can be explained best of all with the absence of direct observation of $\sigma_0$- mesons in experiments (all data about $\sigma_0$- mesons were obtained from the PWA analysis).

However the understanding of properties of $\sigma_0$ - mesons is important not only from the point of view of theory but also for the analysis of data which are obtained at the HADES spectrometer [7] and for an experiment on searching the mixed phase in nuclei-nuclei collisions, which is planned at the nuclotron of the VBLHE JINR [8].

## 2. Selection of kinematics criteria and results of investigations

This paper continues a series of our works, devoted to the study of the $\pi^+\pi^-$-system, with different kinematics criteria [9].

The data were obtained in an exposure of the 1m $H_2$ bubble chamber of LHE (JINR) to a quasimonochromatic neutron beam ($\Delta P_n / P_n \approx 2.5\%$, $\Delta\Omega_{channel} = 10^{-7}\, sterad.$) due to the acceleration of deuterons by synchrophasotron of LHE. The neutron channel has been detailed in [10].

The accuracy of momenta of secondary charged particles from the reaction $np \to np\pi^+\pi^-$ are: $\delta P \approx 2\%$ for protons and $\delta P \approx 3\%$ for $\pi^+$ and $\pi^-$. The angular accuracy was $\leq 0.5°$.

The channels of the reactions were separated by the standard $\chi^2$–method taking into account the corresponding coupling equations [11-13].

In this work in comparison with previous ones we made more hard selection of events taking into account track measurement precise. Border value of $\chi^2$ for using the events from the reaction channel $np \to np\pi^+\pi^-$ is $\chi^2 < 1.5$ (fig.1a). Some cut of missing mass reconstructed after fit-procedure was made.

At fig.1b the missing mass distribution in the selected reaction $np \to np\pi^+\pi^-$ is shown. One can see that the distribution has a maximum at the missing mass equal to the neutron mass with accuracy of 0.1 MeV/c$^2$ and is symmetric about the neutron mass. The width at the half-height is 20 MeV/c$^2$.

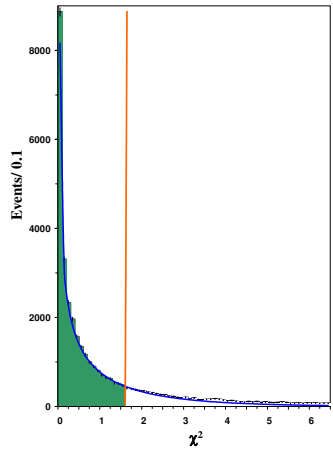 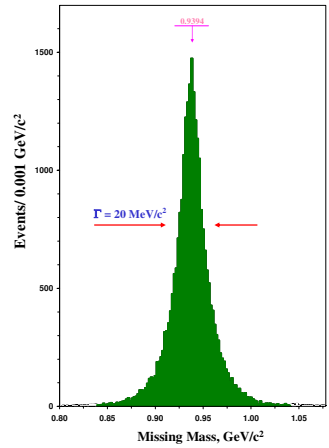

Figure 1a                                                         Figure 1b
The experimental (histogram) and the theoretical (curve)          The missing mass distributions
$\chi^2$–distributions for the reaction $np \to np\pi^+\pi^-$.    for the reaction $np \to np\pi^+\pi^-$.

For more purity of data a small number of events with the missing masses out of green range were excluded.

25864 events of the $np \to np\pi^+\pi^-$ reaction are lost for the processing after application of all cuts. Note that an admixture of other reactions practically is absent.

Earlier, we have already studied the reaction $np \to np\pi^+\pi^-$ [14], and OPE-exchange with a dominated exchange of the charged $\pi$-meson has been shown to be main mechanism of this reaction. It leads to a plentiful production (up to 70% of the total reaction cross section) of $\Delta^{++}$ and $\Delta^-$-resonances in the lower and upper vertices of the corresponding diagrams. The OPE mechanism gives a main part into the events with neutron flying into the forward hemisphere.

Therefore, it seems reasonable to study the resonances in the $\pi^+\pi^-$-system of the reaction $np \to np\pi^+\pi^-$ selecting the events on condition that $\cos\Theta_p^* > 0$. The total contribution of the $\Delta^{++}$ and $\Delta^-$-resonances is no more than 17% for these events, and the background from resonance decays decreases greatly. The number of events with $\cos\Theta_p^* > 0$ is equal to 7647.

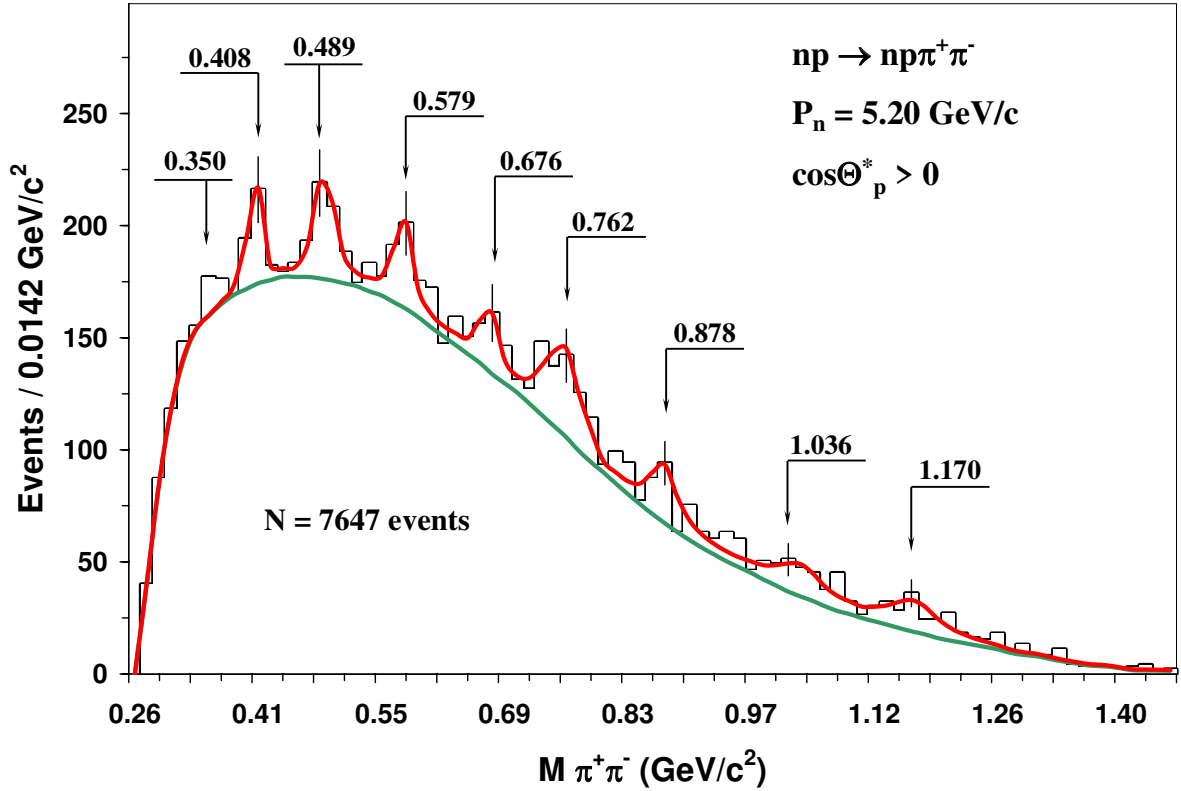

Figure.2

The effective mass distribution of $\pi^+\pi^-$ - combinations
from the reaction $np \to np\pi^+\pi^-$ at $P_n = (5.20 \pm 0.12)$ GeV/c.
Green line – the background curve taken in the form of Legendre polynomial of $9^{th}$ degree.
Red line – the sum of the background curve and the 8 resonance curves taken in the Breight-Wigner form.

Figure.2 shows the effective mass distribution of $\pi^+\pi^-$-combinations for the events with a secondary proton flying into the forward hemisphere in the c.m.s. of the reaction.

The mass distribution is approximated by an incoherent sum of the background curve taken in the form of a superposition of Legendre polynomials and by resonance curves taken in the Breight-Wigner form.

The requirements to the background curve are the following:
- firstly, the errors of the coefficients must be not more than 50 % for each term of the polynomial;
- secondly, the polynomial must describe the experimental distribution after "deletion" of resonance regions with $\overline{\chi^2} = 1.0$ and $\sqrt{D} = 1.41$ (the parameters of $\chi^2$ – distribution with 1 degree of freedom).

The distribution on Fig.2 is approximated by Legendre polynomial of the $9^{th}$ power and by 8 resonance curves. The experimental values of the resonance masses (obtained by fitting procedure) are shown by arrows. The values $\overline{\chi^2}$ and $\sqrt{D}$ for the background curve of the distribution in Fig.2 (with resonance regions excluded) are equal to $\overline{\chi^2} = 0.98 \pm 0.19$ and $\sqrt{D} = 1.47 \pm 0.14$. The contribution of the background to this distribution is 89%.

The same values for the background curve normalized to 100% of events in the plot

(with resonance regions included) are equal to $\overline{\chi^2} = 1.20 \pm 0.15$ and $\sqrt{D} = 1.54 \pm 0.11$ (Confidence Level 12%).

The results of approximation are given in the Table 1 below.

Table 1

|   | $M_{Res} \pm \Delta M_{Res}$ MeV/c² | $\Gamma_{Res}^{exp} \pm \Delta\Gamma_{Res}^{exp}$ MeV/c² | $\Gamma_{Res}^{true} \pm \Delta\Gamma_{Res}^{true}$ MeV/c² | $\sigma_{\mu b}$ | S.D. |
|---|---|---|---|---|---|
| 1 | 408 ± 3 | 11 ± 8 | 7 ± 9 | 12 ± 6 | 3.5 |
| 2 | 489 ± 3 | 20 ± 10 | 16 ± 11 | 20 ± 8 | 4.0 |
| 3 | 579 ± 5 | 17 ± 14 | 7 ± 14 | 18 ± 8 | 3.8 |
| 4 | 676 ± 7 | 11 ± 14 | 16 ± 15 | 11 ± 6 | 3.0 |
| 5 | 762 ± 11 | 53 ± 33 | 48 ± 33 | 26 ± 8 | 6.1 |
| 6 | 878 ± 7 | 30 ± 14 | 11 ± 16 | 11 ± 5 | 3.6 |
| 7 | 1036 ± 13 | 61 ± 30 | 50 ± 33 | 15 ± 5 | 5.1 |
| 8 | 1170 ± 11 | 65 ± 33 | 51 ± 35 | 11 ± 4 | 5.8 |

*The first column* contains the experimental values of the resonance masses (including errors) obtained in the process of approximation.

*The second column* contains the experimental values of the resonance widths.

*The third column* contains the true values of the total width of the resonances. The mass resolution function [15] grows with increasing mass as:

$$\Gamma_{res}(M) = 4.2 \left[ \left( M - \sum_{i=1}^{2} m_i \right) / 0.1 \right] + 2.8, \text{ where:}$$

M – the mass of the resonance, $m_i$ – the rest mass of the particles composing this resonance, M and $m_i$ are in GeV/c²; coefficients 4.2 and 2.8 are in MeV/c².

The true width of a resonance is obtained by formula: $\Gamma_{Res}^{true} = \sqrt{\left(\Gamma_{Res}^{exp}\right)^2 - \left(\Gamma_{res}\right)^2}$.

*The fourth column* contains the production cross sections for the corresponding resonances. For the cross sections errors, we have taken into account the cross section error for the reaction $np \rightarrow np\pi^+\pi^-$ at $P_n = 5.20 \, GeV/c$ ($\sigma_{np \rightarrow np\pi^+\pi^-} = (6.22 \pm 0.28) mb$) [12].

*The fifth column* contains the number of standard deviations of the effects above background: $S.D. = N_{res.} / \sqrt{N_{back.}}$.

The observed resonance at the mass of $M_R = 757 \, MeV/c^2$ has been inserted in RPP since 2000, (S.D. = 6.0) [16].

## 3. Spin and isotopic spin of the resonances

We have tried to estimate the values of spins of the observed resonances in $\pi^+\pi^-$ - system.

To do this, we constructed the distributions of emission angles of $\pi^+$ - meson from the resonance decay with respect to the direction of resonance emission in the general c.m.s. of the reaction. All the values are transformed to the resonance rest system (helicity coordinate system) [17]. The backgrounds are constructed using events at the left and at the right of the corresponding resonance band and subtracted using the weight in proportion to a contribution of a background into resonance region. The result distributions (named "Cos Θ" distribution) are described by a sum of even-power Legendre polynomial (when approximating it was essential that errors in coefficients of the selected Legendre polynomials should be less then 50%) with maximum power being equal to 2J, where J is a spin of the resonance. In such a manner the value of the lower limit of the resonance spin is estimated The authors are grateful to Dr. V. L. Lyuboshitz for the arrangement of the corresponding formulas.

The distributions of this angle are shown in Fig. 3. for the resonances at $M_R = 489\ MeV/c^2$, $M_R = 579\ MeV/c^2$ and $M_R = 762\ MeV/c^2$. The solid line corresponds to the isotropic distribution ($J = 0$).

The distribution for other resonances have the same shape.

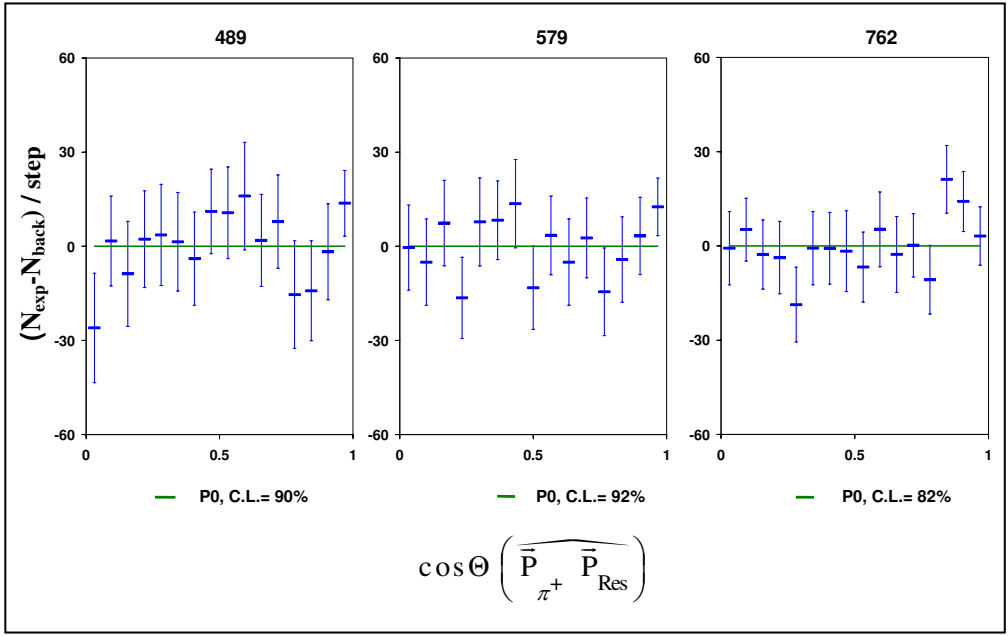

Figure 3
The Cos Θ distributions for the resonance spin estimations.
Masses under investigation and statistical significances of description of corresponding distributions are shown near every graph.

Therefore, the most probable spin values for these resonances are equal to 0.

We don't observe corresponding resonances either in the $\pi^-\pi^0$ - system of the $np \to pp\pi^-\pi^0$ reaction, or in the $\pi^-\pi^-$ - system of the $np \to pp\pi^+\pi^-\pi^-$ reaction, which also have been studied by us [15].

Therefore, it can be affirmed that all resonances which are observed by us have the quantum numbers $I^G(J^{PC}) = 0^+(0^{++})$ and may be identified as $\sigma_0$ - mesons.

## 4. Comparison with other data

A large number of publications are dedicated to the search and study of $\sigma_0$-mesons (see [16]). All of them are based on the PWA of $\pi N$ or $\tilde{p}p$ - interactions. The obtained $\sigma_0$-meson masses ranging from 400 to 1200 $MeV/c^2$ coincide with the sequence of masses observed in our experiment. Especially good coincidence can be obtained in the K-matrix approach [16]. However widths of resonances extracted from PWA are considerably larger than those obtained in our experiment. It may be necessary to use other ideas and more complicated methods of analysis to understand these phenomena better.

## 6. Acknowledgement

The authors are grateful to prof. A.I. Malakhov and prof. S. Vokal for the equipment and organize help to the processing of works at the *np*-collision investigations at the 1m HBC of the VBLHE and for help to the presentation of obtained results.

We also thank to prof. V.L. Lyuboshitz for the persistent help during discussions on physical results.

Our peculiar grateful to the group of operators of the chamber frames processing (the NEOFI subdivision of the VBLHE) for their hard and very professional work in the track preview and measurement.

Also the preview and measuring devices service group headed by V.F. Rubtzov (the LIT of the JINR) is worthy of big grateful.

**References**

1. Ericson T., Weise W. – Pions and Nuclei. Claredon Press, Oxford, 1988
2. Carlson C.E. et al. – Phys. Rev., 1983, v.D27, p.1556;
   Donoghue J.F. et al. – Phys. Lett., 1987, v.B99, p.416
3. Geiger P., Isgur N. - Phys. Rev., 1993, v.D47, p.5050;
   Kokoski R., Isgur N. - Phys. Rev., 1987, v.D35, p.907
4. Nambu Y., Jona-Lasinio G. - Phys. Rev.,1961,v.122,p.355;
   Hatsuda T., Kunihiro T. - Phys. Rep., 1994, v.247, p.22;
   Delbourgo R., Scadron M.D. – Mod. Phys. Lett., 1995, v.A10, p.251;
   Tornquist N.A., Roos M. – Phys. Rev. Lett., 1996, v.76, p.1575;
   Volkov M.K. et al., - JINR, E2-98-101, Dubna, 1998.
5. Bordes J., Lanik J. – Phys. Lett., 1989, v.B223, p.251.
6. Ellis J.,Lanik J. - Phys. Lett., 1985, v.B150, p.289;
   Narison S.–Z. Phys. C, 1984, v.26, p.209
7. A. Kugler -High Acceptance Di-Electron Spectrometer - Tool for Study of Hadron Properties in the Nuclear Medium. Czech Journ. Phys. 50/S2, (2000) 72
8. Sissakian A.N et al. - VIII International Workshop "Relativistic Nuclear Physics: from Hundreds MeV to TeV", Dubna,Russia, May 2005, Proceedings-Dubna, JINR, 2006, p,306
9. Troyan Yu.A. et al. – JINR Rapid Communication, 1998, №5[91] - 98, p.33;
   XIII International Seminar on High Energy Physics Problems «Relativistic Nuclear Physic&Quantum Chromodynamics», Dubna, Russia, Sept.1998.
   Troyan Yu.A. et al. – Particles and Nuclei, Letters, 2000, №6[103] - 2000, p.25;
   XIV International Seminar on High Energy Physics Problems «Relativistic Nuclear Physic&Quantum Chromodynamics», Dubna, Russia, Sept.2000.
   Troyan Yu.A. et al. – Particles and Nuclei, Letters, 2002, №5[114] - 2002,
   XV International Seminar on High Energy Physics Problems «Relativistic Nuclear Physic&Quantum Chromodynamics», Dubna, Russia, Sept.2002.
10. Gasparian A.P.et al. - Pribory i Teknika Eksp., 1977, v.2,p.37; JINR, 1-9111, Dubna, 1975;
11. Moroz V.I.et al. – Yad. Fiz., 1969, v.9, p.565
12. Besliu C.et al. - Yad. Fiz., 1986, v.63, p.888.
13. Yu.A.Troyan et al., Phys.Atom.Nuc., Vol.63, No.9, 2000, pp.1562-1573 [Yad.Fiz., vol.63, No.9. 2000, pp.1648-1659].
14. Ierusalimov A.P. et al. – JINR Rapid Communication, 1989, №2[35] - 89, p.21
15. Troyan Yu.A. et al. – JINR Rapid Communication, 1996, №6[80] - 96, p.73.
16. W.M. Yao et al. (Particle Data Group), - J. Phys. G 33, 1 (2006), http://pdg.lbl.gov
17. Baldin A.M. et al., - Kinematics of Nuclear Reactions (Oxford Univ.Press, London, 1961, transl. of 1$^{st}$ Russ.ed; Atomizdat. M., 1968, 2$^{nd}$ ed.)